\documentclass[twocolumn,twoside,slac_two]{revtex4}
\usepackage{graphicx}
\usepackage{fancyhdr}
\usepackage{overpic,color}
\usepackage{multirow}
\usepackage{bigstrut}
\newcommand{\numutonutau}{$\nu_{\mu}\rightarrow\nu_{\tau}$ }
\newcommand{\numutonue}{$\nu_{\mu}\rightarrow\nu_{e}$ }

\newcommand{\numu}{$\nu_{\mu}$}
\newcommand{\nutau}{$\nu_{\tau}$}
\newcommand{\numubar}{$\overline{\nu}_{\mu}$}
\newcommand{\nue}{$\nu_{e}$}
\newcommand{\nuebar}{$\overline{\nu}_{e}$}

\newcommand{\Uo}[1]{{|U_{#1}|^2}}
\newcommand{\Ut}[2]{{|U_{#1}|^2|U_{#2}|^2}}

\begin{document}

\title{Analysis of Neutral Current Interactions in MINOS: A Search for Sterile Neutrinos}

%

\author{Alexandre Sousa, for the MINOS Collaboration}
\affiliation{Department of Physics, Harvard University, Cambridge, Massachusetts 02138, USA}

\begin{abstract}
A search for disappearance of active neutrinos over a baseline of 735 km was conducted using the NuMI neutrino beam and the MINOS detectors. The data analyzed correspond to an exposure of $3.18\times10^{20}$~protons on target.
The data are fitted to neutrino oscillation models in which mixing with one sterile neutrino is assumed. A comparison of the neutral-current-like spectrum at the FD with the expectation derived from the near detector measurement shows that the fraction of disappearing muon neutrinos converting to a sterile state is less than 52\% at the 90\% confidence level. In addition, the possibility of decay of active neutrinos into sterile species occurring concurrently with neutrino oscillations was analyzed. Pure neutrino decay is disfavored at $5.4\,\sigma$ as an alternate explanation to oscillations for the depletion of muon neutrinos  at 735 km. 
The methodology employed in the analysis of neutral current neutrino events observed in MINOS is described and newly obtained results are presented.
\end{abstract}

\maketitle

\thispagestyle{fancy}


\section{The NuMI Beam and the MINOS Detectors}
The MINOS (Main Injector Neutrino Oscillation Search) experiment is a complete long-baseline neutrino oscillation study. The NuMI (Neutrinos from the Main Injector) neutrino beam created at Fermi National Accelerator Laboratory (Fermilab) is sampled first by the Near detector (ND), on-site at Fermilab, at 1\,km from the target, and then by the Far detector (FD), 735~km away in the Soudan Underground Laboratory in Minnesota. Neutrino oscillation phenomena are studied  by comparing the reconstructed neutrino energy spectra at the Near and Far locations. 

The NuMI neutrino beam is produced using 120\,GeV protons from the Main Injector. The protons are delivered in 10\,$\mu$s spills with up to 4.0$\times 10^{13}$ protons per spill. Positively charged particles produced by the proton beam in a graphite target (mainly $\pi^{+}$ and $K^{+}$) are focused by two pulsed parabolic horns and are then allowed to decay in a 675\,m long, 2\,m diameter decay pipe. The target position relative to the first horn and the horn current are variable.  The data employed in this analysis were obtained using the low energy beam configuration, in which the peak neutrino energy is 3.3\,GeV~\cite{ref:CCPRD}, and were recorded between May 2005 and July 2007, corresponding to a total exposure of 3.18$\times 10^{20}$~protons on target (POT).
The charged current (CC) neutrino event yields at the ND are predicted to be 
91.8\% \numu, 6.9\% \numubar, 1.2\% \nue{} and 0.1\% \nuebar. 

The MINOS detectors are designed to be as similar as possible in order 
to minimize systematic uncertainties. The detectors are fine-grained tracking
calorimeters with an inch thick absorber layer of steel and a 1~cm active 
layer of plastic scintillator constituting one ``plane''. 
Each scintillator layer is constructed from 4.1~cm wide strips. Signals from the scintillator are collected via wavelength-shifting (WLS) fibers and 
carried by clear optical fibers to photomultiplier tubes (PMTs). The ND and FD are both magnetized with a current-carrying coil producing an average field of 1.3\,T in the fiducial volume. 

The 0.98\,kton ND, 103\,m underground, has 282 irregular 4$\times$6\,m$^{2}$ octagonal planes.  The detector consists of two sections, a calorimeter encompassing the upstream 121 planes and a spectrometer containing the downstream 161 planes. In both sections, one out of every five planes is fully covered with 96 scintillator strips attached to the steel plates. In the calorimeter section, the other four out of five planes are partially covered with 64 scintillator strips whereas in the spectrometer section no scintillator is attached to the steel.  The 5.4\,kton FD, 705\,m underground, has 484 octagonal, 8\,m wide instrumented planes. Due to the ND proximity to the target, the signal rate in the ND is $\sim10^{5}$ times larger than in the Far detector.

\section{Neutrino Interactions in the MINOS Detectors}
There are two main types of neutrino interactions observed in the MINOS detectors: Charged Current (CC) interactions, proceeding through the exchange of a $W^{\pm}$ boson with creation of the associated charged lepton, typically defined by a long muon track accompanied by a small hadronic shower at the event vertex caused by nuclear fragmentation; Neutral Current (NC) interactions, proceeding through exchange of a $Z^0$ with the neutrino leaving the detector, which appear as diffuse showers with typical length much shorter than the length of a muon track in a CC event. Correct identification of NC events in the MINOS detectors is difficult due to to short CC events with high hadronic inelasticity, featuring a short muon track often concealed by the hadronic shower or easily confused with the short charged pion tracks also found in NC events. 
\begin{figure} [!h]
\begin{center}
\includegraphics[width=2.5in,height=1.50in]{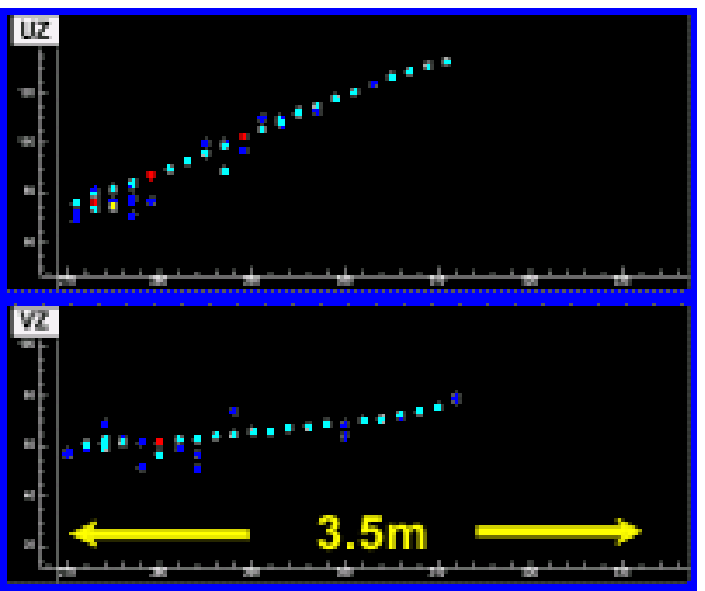}
\includegraphics[width=2.5in,height=1.50in]{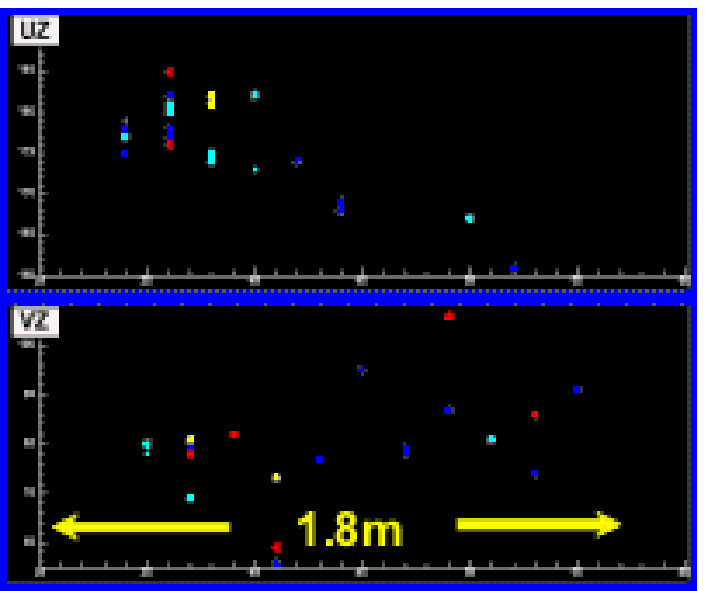}
\end{center}
\caption{\label{fig:ccncfig} 
Example of CC (top) and NC (bottom) interactions in the MINOS detectors. Longitudinal views UZ and VZ are shown in both cases. CC events are characterized by a long muon track and hadronic activity at the vertex, whereas NC events are shorter, displaying a diffuse hadronic shower.
}
\end{figure}

\section{Near Detector Selection}
Events reconstructed in the fiducial volume of the Near and Far detectors are very similar. However, in the ND, an average of 16 neutrino interactions are produced for each 1.8\,$\rm{\mu s}$ spill during typical running with intensities of $2.2\times10^{13}$\,POT per beam spill. Reconstruction algorithms are designed to handle this high interaction rate, but for certain event subclasses shortfalls have been identified and quantified using special studies including low intensity beam data and visual scanning.

\begin{figure}
\begin{center}
  \includegraphics[width=1.05\linewidth]{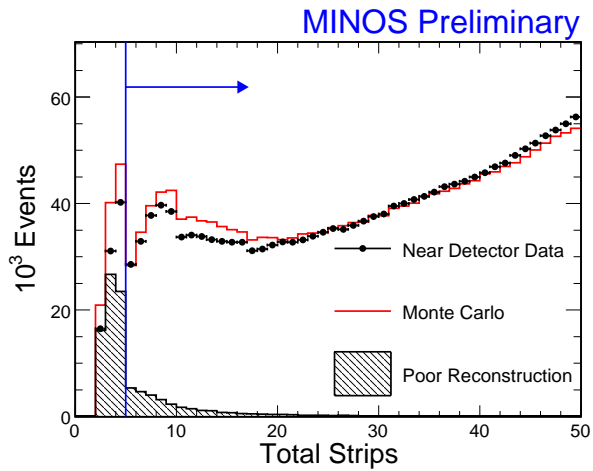}
  \caption{The distribution of the number of strips with non-zero pulse-height, per event, after all other selections are applied. A sizable contribution from poorly reconstructed events (hatched histogram) is observed at low strip counts. The event range accepted for the analysis is identified by the arrow. }
  \label{fig:strips_cleaning}
\end{center}
\end{figure}

Reconstruction failures are classified into three main categories: i) {\it split events}, ii) {\it leakage events}, and iii) {\it incomplete events}. Split events occur when a single neutrino interaction results in two or more reconstructed events. Leakage events are due to incorrectly assigned event vertices causing neutrino interactions outside the fiducial volume to be reconstructed within it. The incomplete event category is a looser classification that refers to further types of failures in shower reconstruction to be described below. In all three categories, the visible energy of a neutrino candidate may be underestimated, resulting in a background to NC events at low energies. As ND data are used to predict the expected spectra at the FD, reconstruction failures specific to the ND must be minimized. A set of selection requirements based on spatial and timing criteria, as well as on activity on the ND sparsely instrumented regions, was developed~\cite{ref:rauferThesis} to reduce the occurrence of these failure modes. The selection criteria can be summarized as follows: i) the time separation between events, $\Delta t$, must exceed 40\,ns; ii) if 40\,ns~$<\,\Delta\,t\,<$~120\,ns, the spatial separation between events, $\Delta z$, must exceed 1\,m; iii) the ratio between the number of active strips per event plane and total number of active planes in the event must be less than unity; iv) for events with less than 5\,GeV of reconstructed energy in which the number of planes is larger in the reconstructed shower than in the reconstructed track, the number of event strips reconstructed  in the detector's veto regions should be less than four, or else the total pulse-height in those regions must be less than 2\,MIP\footnote{Minimum Ionizing Particle, equivalent to the detector response to a perpendicular 1\,GeV-muon traversing one scintillator plane.}; and v) the total number of strips reconstructed in the event must be more than four. Only events that satisfy these criteria are used for further analysis.  After applying these requirements, the background of poorly reconstructed events having visible energy below 1\,GeV has been reduced from 34\% to 8\%. The distribution of the number of occupied strips per event is shown in Fig.~\ref{fig:strips_cleaning}.

\section{Neutral Current Event Classification}
After the selection criteria described above are applied, the analysis proceeds by distinguishing NC events from CC events. The goal of the event classification is to maximize the efficiency and purity of selected samples of NC and CC events, with efficiency defined as the number of true events of one type which are classified as that type,  divided by the total number of true events of that type. Purity is defined as the ratio of the number of true events of one type selected to the total number of events selected as that type. 

\begin{figure}
\begin{center}
\begin{overpic}[width=1.05\linewidth]{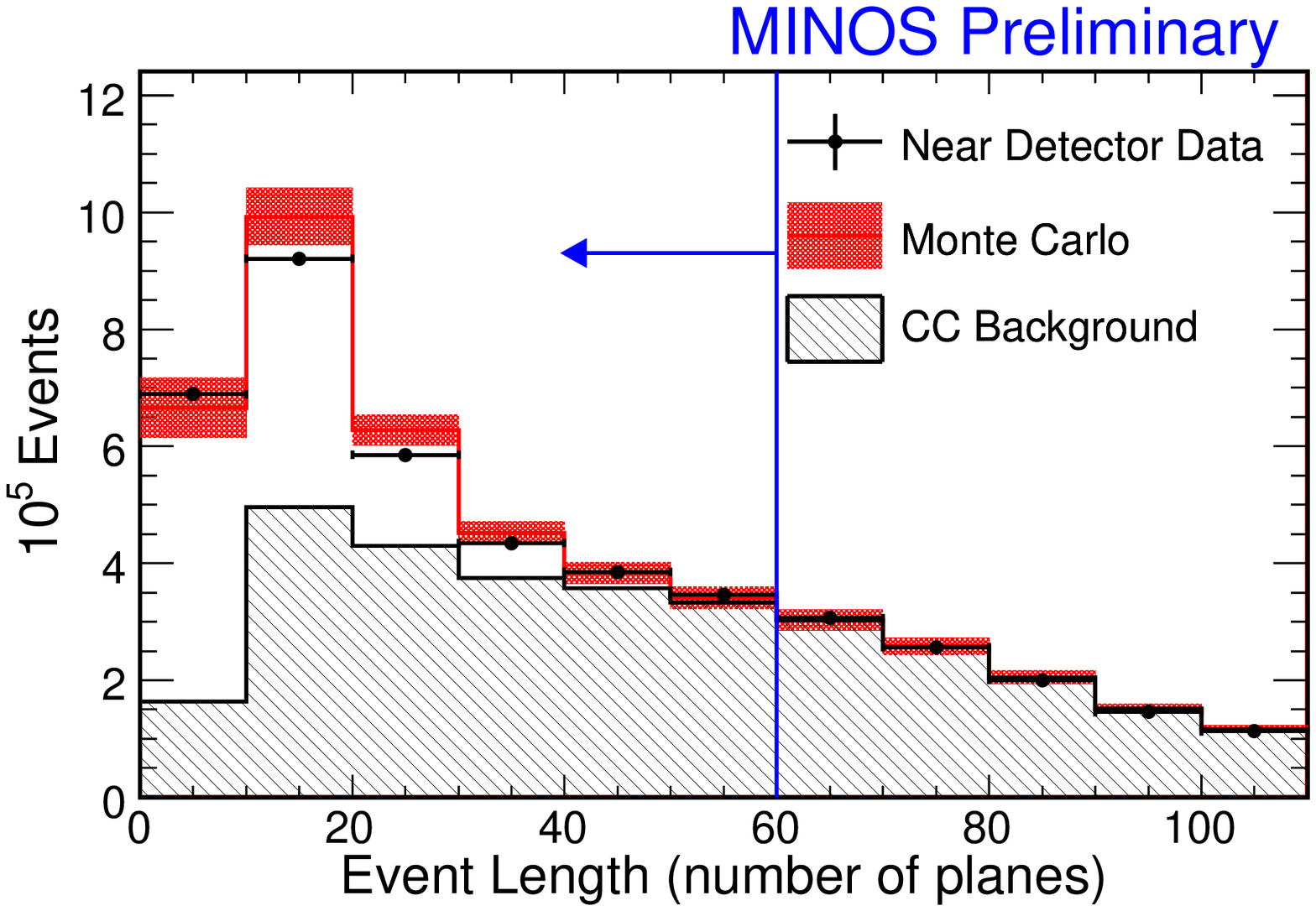}
        \put(12.8,54){\Large\bf a}
\end{overpic}
\begin{overpic}[width=1.05\linewidth]{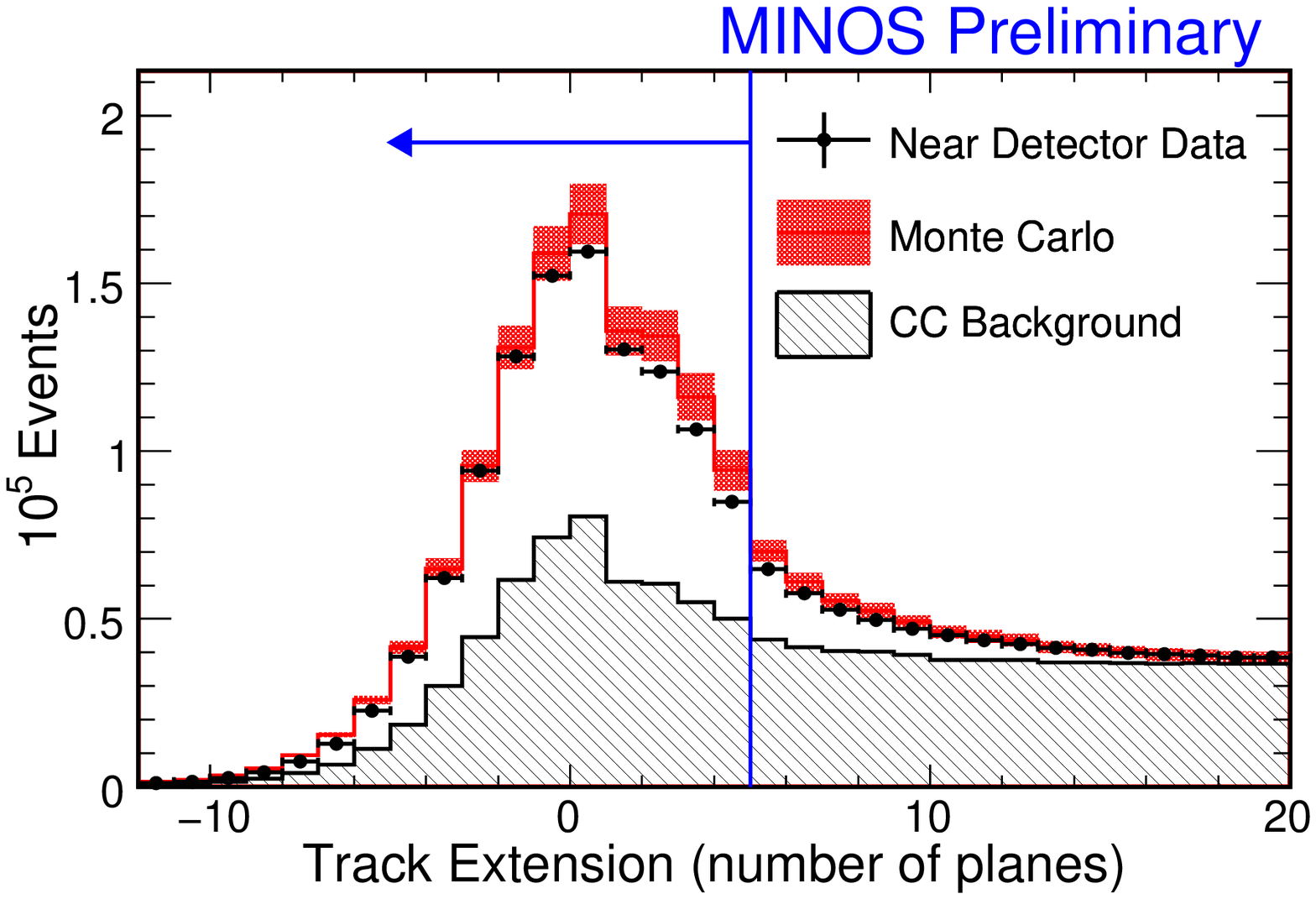}
        \put(12.8,52){\Large\bf b}
\end{overpic}
\caption{Comparisons of ND data with Monte Carlo predictions for distributions of the variables (a) {\it event length} and (b) {\it track extension}. Systematic uncertainties are displayed as shaded bands on the Monte Carlo expectation. Events selected as NC-like are indicated by the arrows. }
\label{fig:NCPID}
\end{center}
\end{figure}

A sample of candidate NC events is obtained by applying specific requirements on three classification variables:  {\it event length}, expressed as the difference between the first and last active plane in the event; {\it number of tracks} reconstructed in the event; and {\it track extension}, defined as the difference between track length and shower length. Events crossing fewer than 60 planes and for which no track is reconstructed are classified as NC. Events crossing fewer than 60 planes that contain a track are classified as NC if the track extends less than 5 planes beyond the shower. The values chosen maximize sensitivity for detection of sterile-neutrino admixture.  Finally, events that are not classified as NC-like are checked to determine if they are CC-like according to classification procedures described in a previous MINOS publication~\cite{ref:CCPRD}.  These requirements are applied to both ND and FD to obtain NC and CC event samples. Distributions for the event length and track extension classification variables for data of the ND are shown in Figs.~\ref{fig:NCPID}a and~\ref{fig:NCPID}b.   The data are plotted together with the prediction of the MINOS Monte Carlo simulation, which adequately reproduces the shapes of the classification-variable distributions.
\section{Near to Far Extrapolation}
The predictions of the energy spectra of the NC and CC samples at the FD are based on the observed ND data and make use of the expected relationship between the neutrino fluxes at the two sites.  The process of making the predictions is called ``extrapolation''  and may be viewed as making corrections to the simulation of interactions in the FD based on the energy spectrum measured in the Near detector.  The current analysis uses an extrapolation technique called the ``Far over Near (F/N)'' method~\cite{ref:CCPRD,ref:KoskinenThesis}. This method makes the prediction of the FD spectrum by taking the product of two quantities: The ratio of the expected number of events from the Monte Carlo simulation for each energy bin in the FD and ND spectra; and the number of observed ND data events. The F/N method prediction is robust against distortions arising from differences between data and Monte Carlo simulation in the ND as these distortions are translated to the FD and do not affect the oscillation measurement~\cite{ref:CCPRD}. 
For instance,  for the case of the \numu~CC component of the NC and CC
samples, the F/N extrapolation predicts the number of events at the FD for the $i$-th bin of reconstructed energy to be
\begin{equation}
  \displaystyle
  F_{i}^{predict}=N_{i}^{data}\left(~\frac{\displaystyle
  \sum_{j}F_{ij}^{MC}~P_{\nu_\mu\rightarrow\nu_{x}}(E_j)}{N_{i}^{MC}}\right),
  \label{eq:FoverN}
\end{equation}
where $N_{i}^{data}$ is the number of selected events in the $i$-th reconstructed energy bin in the ND and $N_{i}^{MC}$ is the number of events expected in that bin from the ND Monte Carlo simulation. The $F_{ij}^{MC}$ represents the number of events expected from the FD Monte Carlo simulation in the $i$-th bin of reconstructed energy and $j$-th bin of true neutrino energy. In the equation, $E_{j}$ is the true neutrino energy and $P_{\nu_\mu\rightarrow\nu_{x}}$ the probability of muon neutrino transition to any other flavor.

\section{Systematic Uncertainties}
The main sources of systematic uncertainties in the analysis are: i) {\it absolute scale of the hadronic energy}; ii) {\it relative calibration of the hadronic energy} in the two detectors; iii) {\it relative normalization} between the two detectors; iv) {\it CC background} in selected NC events; and v) {\it uncertainties due to the ND selection requirements} in the ND event counts. Monte Carlo studies have been performed where for each single uncertainty the Monte Carlo spectrum is varied by $\pm$1 standard deviation independently, in order to estimate the effect of each on the extrapolated spectrum. Beam and cross-section uncertainties that are common to the two detectors effectively cancel when using the F/N extrapolation.

The absolute hadronic energy scale has an uncertainty of 12\%.  This value is a combination of the uncertainty in the hadronization model and intranuclear effects (10\%) and uncertainty of the detector response to single hadrons (6\%).  The relative energy scale between the two detectors has an uncertainty of 3\%. The relative normalization between the two detectors has an uncertainty of 4\%. This is a combination of the uncertainties due to fiducial mass, live time, and reconstruction differences between the two detectors. 

To evaluate the uncertainties due to the ND selection, the requirement that the total number of reconstructed strips in an event is at least four was shifted by $\pm1$ strip. The effects on the reconstructed energy spectrum were determined for each shift. The uncertainty has been estimated to be 15.2\% for $E_{\text{reco}}< $0.5\,GeV; 2.9\% for 0.5\,GeV~$< E_{\text{reco}}<$~1.0\,GeV; 0.4\% for 1.0\,GeV$< E_{\text{reco}} <$~1.5\,GeV and is negligible fpr higher visible energies.

Finally, the uncertainty in the number of CC background events is determined using ND data taken in several different beam configurations. Using the observed differences in energy spectrum between the low-energy beam configuration and each of the other beam configurations along with information from Monte Carlo simulation of each configuration, the uncertainty on the CC background number is found to be 15\% for all energy ranges at the Near and Far detectors~\cite{ref:NCPRL}.

\section{Search for Sterile Neutrinos}
Results from the MiniBooNE experiment\,\cite{ref:MiniBoone}, which looked for \numutonue or $\overline{\nu}_\mu\rightarrow\overline{\nu}_e$ appearance on a short baseline of 540\,$\mathrm{m}$ using a neutrino beam with a mean energy of $\sim 700$\,$\mathrm{MeV}$, mostly rule out the potential scenario of oscillations at the mass-squared $1\,\mathrm{eV}^2$ scale, a possibility reported by the LSND experiment\,\cite{ref:lsnd}. Scenarios including active-sterile neutrino mixing with one or more sterile neutrinos have been put forward to explain the LSND result. Such scenarios, for which sterile neutrino oscillations occur over short baselines, are now essentially ruled out as an explanation for LSND~\cite{ref:Maltoni2007}, but the possibility of sterile mixing over long baselines remains. MINOS searches for sterile neutrino oscillations in the atmospheric sector, in a long baseline setting, by studying disappearance of NC events measured at the FD relative to the flux observed at the ND. NC events are not affected by \numutonutau oscillations, but would suffer an energy-dependent depletion if $\nu_{\mu}\rightarrow\nu_{s}$ oscillations were to occur.

\subsection{Active Neutrino Disappearance}
\begin{figure}
  \includegraphics[width=1.05\linewidth]{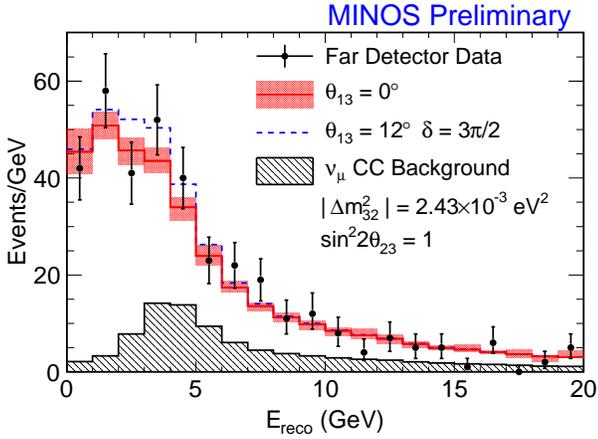}
  \caption{The reconstructed energy spectrum of NC selected events at the FD (points with statistical uncertainties). The Monte Carlo prediction assuming standard three-flavor oscillations is also shown, both with (dashed line) and without (solid line) $\nu_e$ appearance at the CHOOZ limit. The shaded region indicates the 1 standard deviation systematic uncertainty on the prediction. The hatched region shows the Monte Carlo prediction for the background of misidentified CC events in this sample.}
  \label{fig:fd_spectrum}
\end{figure}

A three neutrino analysis assuming oscillations occur only among active flavors is carried out to search for active neutrino disappearance. 
A total of 388 data events are selected as NC in the Far detector.  The measured and predicted $E_{\text{reco}}$ spectra at the FD are shown in Fig.~\ref{fig:fd_spectrum}.  The prediction assumes oscillations with the values of $|\Delta m^{2}_{32}|$ and $\theta_{23}$ previously measured by MINOS~\cite{ref:CCPRL}.
Although this analysis is not capable of isolating an electron neutrino appearance signal, it must take $\nu_\mu\rightarrow\nu_e$ oscillations into account because the classification criteria of this analysis include \nue~CC interactions in the NC enriched sample with nearly 100\% efficiency. This is done by comparing the observed NC spectrum to two predictions, one that assumes null \nue~appearance, and another that assumes an upper limit for the \nue~appearance rate in the FD calculated with the normal neutrino-mass hierarchy, $\theta_{13} = 12^\circ$, and $\delta = 3\pi/2$.  The choice of $\theta_{13}$ corresponds to the 90\% confidence level upper limit established by the CHOOZ reactor experiment~\cite{Apollonio:2002gd} for the $|\Delta m^{2}_{32}|$ value measured by MINOS~\cite{ref:CCPRL}.  As seen in Fig.~\ref{fig:fd_spectrum}, the observed spectrum matches the prediction based on oscillations among the three active flavors quite well over the full range of allowed values of $\theta_{13}$.

The agreement between the observed and predicted NC spectra is quantified using a statistic, $R$:  
\begin{equation}
R \equiv \frac{N_{\text{Data}}-B_{\text{CC}}}{S_{\text{NC}}},
\label{eq:fdef}
\end{equation}
where, within a given energy range, $N_{\text{Data}}$ is the observed event count, $B_{\text{CC}}$ is the extrapolated CC background from all flavors, and $S_{\text{NC}}$ is the extrapolated number of NC interactions~\cite{ref:NCPRL}.  Because the disappearance of $\nu_\mu$ occurs mainly for true neutrino energies $< 6$~GeV~\cite{ref:CCPRL}, the data are separated into two samples.  Events with $E_{\text{reco}}<3$~GeV are grouped into a low-energy sample while events with $3~\text{GeV}<E_{\text{reco}}<120$~GeV are grouped into a high-energy sample. The values of $S_{\text{NC}}$ and contributions to $B_{\text{CC}}$ for the two energy ranges are shown in Table~\ref{table:nums}.
\begin{table}
\caption{ \label{table:nums} Values of $R, N_{\text{Data}}, S_{\text{NC}},$ and the contributions to $B_{\text{CC}}$ for various reconstructed energy ranges.  The numbers in parentheses are calculated including \nue~appearance at the CHOOZ limit, as discussed in the text.  The first uncertainty in the value of $R$ shown is the statistical uncertainty, the second is systematic uncertainty, and the third is due to possible \nue~appearance.}  
\begin{tabular}{llcccc}
\hline
$E_{\text{reco}}$ (GeV)      & $N_{\text{Data}}$ & ~~$S_{\text{NC}}$~~ & ~~$B^{\nu_\mu}_{\text{CC}}$~~ & ~~$B^{\nu_\tau}_{\text{CC}}$~~ & ~~$B^{\nu_e}_{\text{CC}}$~~ \\ \hline
$0-3$       &141&125.1&13.3&1.4&2.3~~(12.4)   \\
$3-120$  &247&130.4&84.0&4.9&16.0 (32.8) \\
\hline 
$0-3$ &\multicolumn{5}{l}{$R=0.99\pm0.09\pm0.07-0.08$} \\
$3-120$ &\multicolumn{5}{l}{$R=1.09\pm0.12\pm0.10-0.13$} \\
$0-120$ &\multicolumn{5}{l}{$R=1.04\pm0.08\pm0.07-0.10$} \\
\hline 
\end{tabular}
\end{table}
The measured values of $R$ for each energy range indicate that neutrino oscillations among the active flavors describe well the observed data.  Over the full energy range, $0-120$~GeV, a value of $R=1.04\pm0.08\text{(stat.)}\pm0.07\text{(syst.)}-0.10(\nu_e)$ is measured, corresponding to a depletion of the total NC event rate assuming null (maximally-allowed) \nue~appearance of less than $8\%~($18\%) at 90\% confidence level.
\subsection{Active-Sterile Neutrino Mixing}
The description of mixing between three active neutrino flavors and one sterile neutrino requires the addition of one mass eigenstate, The choice of parameterization for the expanded mixing matrix follows considerations in Ref.~\cite{Donini:2007yf}. Two cases are considered for the placement of $\nu_4$ in the neutrino mass spectrum: {\it i)} $m_{4} = m_{1}$; and {\it ii)} $m_{4} \gg m_{3}$.  In the first case, the first and fourth mass eigenstate are assumed to be degenerate, as are the second and fourth mass eigenstates. These degeneracies imply that $\theta_{14}=\theta_{24}=0^{\circ}$ and the corresponding oscillation probabilities are given by:
\begin{eqnarray}
\nonumber
P_{\nu_\mu\rightarrow\nu_\mu} & = &1 - 4\Uo{\mu 3}\biggl(1-\Uo{\mu 3}\biggr)\sin^{2}\Delta_{31},  \label{eq:oscprobs41} \\
P_{\nu_\mu\rightarrow\nu_{\alpha}}  & = & 4\Ut{\mu 3}{\alpha 3}\sin^{2}\Delta_{31}\biggr.   \\ \nonumber
\end{eqnarray}
where $\alpha = e, \tau, \text{ or } s$ and  $\Delta_{31} \equiv (m^{2}_{3}-m^{2}_{1})L/(4E)$. In the  $m_{4} \gg m_{3}$ case, the fourth mass eigenstate is assumed to be much larger than the third. Consequently the values of $\sin^{2}\Delta _{41}$ and $\sin^{2}\Delta_{43}$ average to $\frac{1}{2}$.  Additionally, $\sin 2\Delta_{41}$ and $\sin 2\Delta _{43}$ average to 0.  In this model $\Delta m^{2}_{43}$ is assumed to be $\mathcal{O}$(eV$^{2}$) such that the regime of rapid oscillations and thus the averages mentioned are valid at the Far site, while ensuring no observable depletion of $\nu_\mu$ occurs at the Near detector. Using the above simplifications, the oscillation probabilities are:
\begin{eqnarray}
\nonumber
P_{\nu_\mu\rightarrow\nu_\mu} & = &1 - 4\biggl\{\Uo{\mu 3}\biggl(1-\Uo{\mu 3}-\Uo{\mu 4}\biggr)\sin^{2}\Delta_{31} \nonumber \\
& & + \frac{\Uo{\mu 4}}{2}(1-\Uo{\mu 4})\biggr\}, \label{eq:oscprobs43} \nonumber \\
P_{\nu_\mu\rightarrow\nu_\alpha}  & = & 4\mathcal{R}\biggl\{\biggl(\Ut{\mu 3}{\alpha 3}+U^{*}_{\mu 4}U_{\alpha 4}U_{\mu 3}U^{*}_{\alpha 3}\biggr)\sin^{2}\Delta_{31} \nonumber \\
& &  + \frac{\Ut{\mu 4}{\alpha 4}}{2} \biggr\},
\end{eqnarray}

The data are compared to Monte Carlo predictions based on the probabilities in Eqs.~(\ref{eq:oscprobs41})~and~(\ref{eq:oscprobs43}) using a $\chi^{2}$ statistic including nuisance parameters for the five systematic uncertainties described above. The best fit values for the mixing angles in the two models as well as the $\chi^{2}$ for each are shown in Table~\ref{tab:fit_pts}.
\begin{table}[h]
  \caption{  \label{tab:fit_pts} Best fit points and uncertainty ranges obtained for the active-sterile oscillation models. Results are shown with and without \nue~appearance at the CHOOZ limit.  All angles are given in degrees.}
\begin{tabular}{|c|*{5}{|c}|}
\hline
    Model & $\theta_{13}$   \bigstrut & $\chi^2$/D.O.F. & $\theta_{23}$  & $\theta_{24}$ & $\theta_{34}$ \\
    \hline\hline
    \multirow{2}{*}{$m_4 = m_1$} & $0$   \bigstrut & 47.5/39 & $45.0^{+9.0}_{-8.9}$ & - & $0.1^{+28.7}_{-0.1}$ \\
                                     & $12$ \bigstrut & 46.2/39 & $47.1^{+8.8}_{-11.0}$ & - & $23.0^{+22.6}_{-24.1}$ \\
    \hline
    \multirow{2}{*}{$m_4\gg m_3$} & $0$   \bigstrut & 47.5/38 & ${45.0}^{+9.0}_{-8.9}$ & ${0.0}^{+7.2}_{-0.0}$ & ${0.1}^{+28.7}_{-0.1}$ \\
                                 & $12$ \bigstrut & 46.2/38 & ${47.1}^{+8.8}_{-11.0}$ & ${0.0}^{+7.2}_{-0.0}$ & ${23.0}^{+22.6}_{-24.1}$ \\
    \hline
  \end{tabular}
\end{table} 

From these results, a limit can be set on the mixing angles, $\theta_{34} < 38^\circ\,(56^\circ)$ at 90\% confidence level for the $m_{4} = m_{1}$ model. The number in parentheses represents the 90\% C.L. limit obtained when maximal $\nu_e$ appearance is allowed. For the $m_{4}\gg m_{3}$ model, $\theta_{24} < 10^\circ\,(11^\circ)$ and $\theta_{34} < 38^\circ\,(56^\circ)$ at the 90\% confidence level.

A perhaps more straightforward way to quantify the coupling between the active and sterile neutrinos is to determine the fraction of disappearing $\nu_\mu$ that transition to $\nu_s$.  That fraction is expressed as
\begin{equation}
 f_{s}\equiv\frac{P_{\nu_\mu\rightarrow\nu_s}}{1-P_{\nu_\mu\rightarrow\nu_\mu}}.
\label{eq:fsdef}
\end{equation}
The 90\% confidence level limit for $f_{s}$ is determined by selecting a large number of test values of $\theta_{23}$ and $\theta_{34}$ from Gaussian distributions with mean and $\sigma$ given in Table~\ref{tab:fit_pts}.  The value of $f_{s}$ that is larger than 90\% of the test cases represents the limit.  For both models, the value corresponding to the 90\% confidence level is  $f_{s} < 0.52\,(0.55)$, with the value in parentheses indicating the value obtained for maximally-allowed $\nu_e$ appearance in the beam. Therefore, approximately 50\% of the disappearing $\nu_\mu$ can convert to $\nu_s$ at 90\% confidence level as long as the amount of $\nu_e$ appearance is less than the limit presented by the CHOOZ collaboration.
\section{Neutrino Oscillations with Decay}
Neutrino decay, as an alternative or companion process to neutrino oscillations offers some capability for reproducing neutrino disappearance trends~\cite{ref:Decay}. The model investigated here~\cite{ref:DecayMaltoni} includes neutrino oscillations occurring in parallel with neutrino decay. In this model, the survival and decay probabilities are:
\begin{eqnarray}
P_{\mu\mu} & = & \cos^4\theta+\sin^4\theta e^{-\frac{m_3L}{\tau_3E}}+\nonumber\\
          &   & 2\cos^2\theta\sin^2\theta e^{-\frac{m_3L}{2\tau_3E}}\cos\left(\frac{\Delta m^2_{32}L}{2E}\right) \nonumber\\
P_{\rm decay} & = & \left(1-e^{-\frac{m_3L}{\tau_3E}}\right)\sin^2\theta. 
\label{eq:p_decay}
\end{eqnarray}
where $\tau_3$ is the lifetime of the $\nu_3$ mass state and $\theta$ is the mixing angle governing oscillations between \numu~and \nutau.
The limits $\tau_3\to\infty$ and $\Delta m^2_{32}\to0$ correspond to a pure oscillations or a pure decay scenario, respectively.
The best fit values extracted  for $\theta$ and the parameter $\alpha\equiv m_3/\tau_3$ using this model are summarized in Table~\ref{tbl:fit_decay}.
\begin{table}
\caption{  \label{tbl:fit_decay} Best fit points and uncertainty ranges obtained for the relevant parameters of the oscillation with decay model. The result obtained for the pure decay scenario, $\Delta m^2_{32}~\rightarrow0$, is also presented. Angles are shown in degrees.}
\begin{tabular}{|c|*{3}{|c}|}
\hline
   Model & $\chi^2$/D.O.F. \bigstrut & $\alpha$~(GeV/km) & $\theta$\\
    \hline
    \hline
    Osc. with Decay \bigstrut & 47.5/39 & $0.00^{+0.90}_{-0.0}\times10^{-3}$ & ${45.0}^{+10.83}_{-8.96}$    \\
    \hline
    Pure Decay  \bigstrut & 76.4/40 & $4.6^{+3.1}_{-2.3}\times10^{-3}$ & ${50.9}^{+39.1}_{-11.27}$ \\
    \hline
  \end{tabular}
\end{table} The results are consistent with maximal mixing ($\theta=45^\circ$) and with no neutrino decay ($\alpha=0$).  The 90\% C.L. limit found for the neutrino decay lifetime is $\tau_{3}/m_{3}~>~2.1\times10^{-12}$\,s/eV.

Both the NC and CC FD spectra are included in the fits, therefore additional sensitivity is gained with respect to previous MINOS analyses of neutrino decay, which used the CC FD spectrum only, since any neutrino decay would also deplete the NC spectrum. A $\Delta\chi^2$ of 28.9 is obtained for the pure decay scenario. Thus, a pure neutrino decay model with null oscillations, as considered in Ref.~\cite{ref:CCPRL}, is disfavored at the level of 5.4 standard deviations, an improvement of 1.7 standard deviations on the previously published value.

\section{Outlook}
The analysis described here utilizes a exposure of 3.18$\times10^{20}$~POT. However, MINOS has already accumulated more than 7$\times10^{20}$~POT of NuMI beam data. An analysis of neutral current events in the additional data is underway with expected improvements of the limits on mixing angles and of the sensitivity to sterile fraction. Updated results are expected to be available in early 2010.
\begin{acknowledgments}
This work was supported by the U.S. DOE, the U.S. NSF, the U.K. STFC, the State and University of Minnesota, the University of Athens, Greece, by FAPESP and by CNPq in Brazil.
\end{acknowledgments}

\bigskip 

\begin{thebibliography}{99}   

\bibitem{ref:CCPRD} P.~Adamson et al.
(\bibinfo{collaboration}{MINOS}), Phys. Rev. D {\bf 77} 072002 (2008); D.~G.~Michael et al., 
(\bibinfo{collaboration}{MINOS}), Phys. Rev. Lett. {\bf 97}, 191801 (2006)
 
\bibitem[{Raufer}(2007)]{ref:rauferThesis}
\bibinfo{author}{{T.~M.} {Raufer}},
\bibinfo{journal}{Ph.D. Thesis, Oxford University}  (\bibinfo{year}{2007}).

\bibitem[{{Koskinen}(2009)}]{ref:KoskinenThesis}
\bibinfo{author}{{D.~J.} {Koskinen}},
  \bibinfo{journal}{Ph.D. Thesis, University College London}
  (\bibinfo{year}{2009}).

\bibitem[{Adamson et~al.}(2008)]{ref:NCPRL}
\bibinfo{author}{{P.}~{Adamson}} {\it et~al.}
  (\bibinfo{collaboration}{MINOS}), 
  \bibinfo{journal}{Phys. Rev. Lett.}
  \textbf{\bibinfo{volume}{101}}, \bibinfo{pages}{221804}
  (\bibinfo{year}{2008}).

\bibitem[{Aguilar-Arevalo et~al.}(2009)]{ref:MiniBoone}
\bibinfo{author}{{A.}~{Aguilar-Arevalo} {\it et~al.}},
(\bibinfo{collaboration}{MiniBooNE}),
\bibinfo{journal}{Phys. Rev. Lett.} \textbf{\bibinfo{volume}{103}},
\bibinfo{pages}{061802} (\bibinfo{year}{2009}).

\bibitem[{Aguilar et~al.}(2001)]{ref:lsnd}
\bibinfo{author}{{A.}~{Aguilar} {\it et~al.}},
(\bibinfo{collaboration}{LSND}),
\bibinfo{journal}{Phys. Rev. D} \textbf{\bibinfo{volume}{64}},
\bibinfo{pages}{112007} (\bibinfo{year}{2001}).

\bibitem[{Maltoni et~al.}(2007)]{ref:Maltoni2007}
\bibinfo{author}{{M.}~{Maltoni}},
\bibinfo{journal}{J. Phys. Conf. Ser.} \textbf{\bibinfo{volume}{110}},
\bibinfo{pages}{082011} (\bibinfo{year}{2008}).

\bibitem[{{Adamson et~al.}(2008)}]{ref:CCPRL}
\bibinfo{author}{{P.}~{Adamson}} {\it et~al.}
  (\bibinfo{collaboration}{MINOS}), \bibinfo{journal}{Phys. Rev. Lett.}
  \textbf{\bibinfo{volume}{101}}, \bibinfo{pages}{131802}
  (\bibinfo{year}{2008}).

\bibitem[{{Apollonio et~al.}(2003)}]{Apollonio:2002gd}
\bibinfo{author}{{M.}~{Apollonio}} {\it et~al.} 
(\bibinfo{collaboration}{CHOOZ}), \bibinfo{journal}{Eur.
  Phys. J. C} \textbf{\bibinfo{volume}{27}}, \bibinfo{pages}{331}
  (\bibinfo{year}{2003}).

\bibitem[{{Donini et~al.}(2007){Donini, Maltoni,
  Meloni, Migliozzi, and Terranova}}]{Donini:2007yf}
\bibinfo{author}{{A.}~{Donini}} {\it et~al.},
   \bibinfo{journal}{J. High En. Phys.} \textbf{\bibinfo{volume}{12}},
  \bibinfo{pages}{013} (\bibinfo{year}{2007}).

\bibitem[{Barger et~al.}(1999)]{ref:Decay}
\bibinfo{author}{{V.}~{Barger} {\it et~al.}},
\bibinfo{journal}{Phys. Rev. Lett.} \textbf{\bibinfo{volume}{82}},
\bibinfo{pages}{2640} (\bibinfo{year}{1999}).

\bibitem[{{Gonzalez-Garcia and Maltoni}(2008)}]{ref:DecayMaltoni}
\bibinfo{author}{{M.~C.} {Gonzalez-Garcia}}
  {and} \bibinfo{author}{{M.}~{Maltoni}},
  \bibinfo{journal}{Phys. Lett. B} \textbf{\bibinfo{volume}{663}},
  \bibinfo{pages}{405} (\bibinfo{year}{2008}).

\end{thebibliography}

\end{document}